\documentstyle[preprint,aps]{revtex}
\begin{document}

\draft
\preprint{\vbox{\hbox{U. of Iowa preprint 96-11}}}

\title{A Two-Parameter Recursion Formula For Scalar Field Theory}

\author{Y. Meurice and G. Ordaz}

\address{Department of Physics and Astronomy\\
University of Iowa, Iowa City, Iowa 52246, USA}

\maketitle

\begin{abstract}
We present a two-parameter family of recursion formulas for scalar field
theory. The first parameter is the dimension $(D)$. The second parameter 
($\zeta$)
allows one to continuously extrapolate between Wilson's approximate recursion
formula and the recursion formula of Dyson's hierarchical model.
We show numerically that at fixed $D$,
the critical exponent $\gamma $ depends continuously on $\zeta$.
We suggest the use of the $\zeta -$independence as a guide to construct 
improved recursion formulas. 
\end{abstract}

\newpage

The renormalization group\cite{wilson,wilson2} 
method has shed light on the related questions
of second order phase transitions and the infinite cut-off limit
of field theories. The practical application of the method usually requires
approximations. In the case of scalar field theory, a particularly
simple approximate renormalization group transformation\cite{wilson} 
is Wilson's 
``approximate recursion formula'' (WARF in the following). 
The WARF is a simple integral equation with only one
variable and has a free parameter which can be adjusted
to approximate a $D$-dimensional theory. 
The WARF can be handled very easily using numerical methods
or various perturbative expansions. 
The numerical value of the critical exponent $\gamma $ 
obtained\cite{wilson} with the the WARF is 1.218 for $D=3$. The WARF 
can also be used to study
non-perturbatively the cut-off dependence of the renormalized quantities.
In the case $D=4$, this can be used to set triviality bounds\cite{hazenfratz} 
on the mass of a scalar particle.

The derivation of the WARF is a masterpiece of quantum calculation.
Unfortunately, it is not based on an expansion in a small parameter
and there no obvious way to restore order by order the details erased
by the approximation. Clearly, one needs an organizing principle to 
improve the WARF. A group theoretical approach\cite{marseille} was proposed
in the case of theories with quadratic interactions but
this approach fails to control the proliferation of non-local
terms in the case of quartic interactions. At the end of this letter,
a new method of improvement will be suggested.
 
A renormalization group transformation ``very closely connected''\cite{gk}
to the WARF holds exactly for Dyson's hierarchical 
model\cite{dyson,baker,sinai}.
In the following, we use DHMRF as short for ``Dyson's hierarchical model
recursion formula''.
The DHMRF is also an integral equation with one variable, and it also has
a free parameter expressible in terms of $D$. 
For definiteness, the WARF and DHMRF are given below by Eqs. (7-8) with
$\zeta =$ 1 and $1/D$ respectively.
Roughly speaking\cite{baker},
the WARF does in one step what the DHMRF does in $D$ steps. However, this
is not an exact statement because\cite{epsi} the value of 
$\gamma $ in $D=3$ is 1.300.
Consequently, the two models have different physical properties.

Before going further, we would like to emphasize that the difference between 
the two values of $\gamma $ quoted above is significantly larger
than the errors involved in the numerical calculations. We have repeated
Wilson's calculation with smaller integration steps, cutting the integral
at larger values and using different criteria to determine criticality.
We found that these changes affected $\gamma $ by less than 0.002.
We also used Wilson's numerical integration method for the DHRMF and found
a value of 1.301 for $\gamma $. We confirmed\cite{prl} 
this result with errors
smaller than 0.003 using 
the first 800 coefficients of the high-temperature expansion. 
Clearly, the difference between 1.22 and 1.30 is more than
25 times larger than the numerical errors involved in each calculation. 

The WARF and the DHMRF can be seen as two approximate versions of the 
(much more complicated) renormalization group transformation for 
a scalar lattice model with nearest neighbor iterations. The WARF
integrates $2^D$ field variables at a time keeping their sum constant
while the DHMRF does the same thing but with only two field variables.
The fact that different values of $\gamma $ are obtained with the two
approximate methods sets a limit on the accuracy of the approximation.
For comparison, the universal value of $\gamma $ in $D=3$ for nearest
neighbor models is approximately 1.25. 

Ideally, one would like to construct a renormalization procedure
for the  nearest neighbor models 
corresponding to the integration of 
\begin{equation}
a \equiv 2^{\zeta D}
\end{equation}
field variables 
at a time and have all the physical quantities independent of $\zeta $.
The quantity $2^{\zeta}$ is a scale factor which would play a role
similar to the arbitrary scale parameter $\mu $ used 
in some version\cite{gross}
of the Callan-Symanzik equations and where
one obtains homogeneous differential equations
of the form ${d\over {d\mu }}$
(physical quantity)=0.
In this letter, we make a first step in a similar direction, by introducing 
a recursion formula where $\zeta $ is arbitrary and which interpolates
continuously between the WARF ($\zeta =1$) and the DHMRF ($\zeta = 1/D$).

This recursion formula can be constructed using 
Dyson's hierarchical model (DHM)
as an explicit realization and then extending the
results for arbitrary $\zeta$.
In other words, this paragraph should first be read  
with $\zeta $ taking the fixed value $1/D$ and 
considered as well-known results written in a slightly unusual way.
We call the sum of the fields in a cube containing $a^l$ sites
$\phi_l$, with $l=1,2..$ corresponding to successive
 renormalization
group transformations.
We recall that $a$ is defined in Eq.(1) and takes
the value 2 for DHM. At criticality, one has the scaling law\cite{parisi} 
\begin{equation}
<(\phi_l)^2>\ \propto (ab)^{2l}
\end{equation} 
where 
\begin{equation}
b\equiv 2^{-{{D-2}\over 2}\zeta} \ .
\end{equation} 
Calling $P_l(x)dx$ the probability for $(\phi _l /(ab)^l)$ to
take a value between $x$ and $x+dx$, we see that at criticality,
\begin{equation}
\rho_l=\int dx P_l(x)x^2 
\end{equation} 
tends to a constant for large $l$.
On the other hand, in the high-temperature phase, 
\begin{equation}
<(\phi_l)^2>\ \propto (a)^{l}
\end{equation}  
and $\rho_l \propto (ab^2)^{-l }$.
$P_l(x)$ is the main object studied with the renormalization group method.
Using the parametrization,
\begin{equation}
P_l(x)\ =\ Ke^{-(1/a)Q_l(b x)},
\end{equation} 
one can check that the recursion formula
\begin{equation}
Q_{l+1}(x)=-a\ ln\lbrack {I_l(bx)\over I_l(0)}\rbrack 
\end{equation} 
with
\begin{equation}
I_{l}(x)=\int_{-\infty}^{\infty}dy e^{-y^2 -{1\over 2} Q_l(x+y)
-{1\over 2} Q_l(x-y)} 
\end{equation} 
is equivalent to the DHMRF. The equivalence 
with the formulation usually found in the literature\cite{sinai} is made clear by 
writing the recursion formula for the quantity $\psi _l(x) = I_{l-1} (bx)$.
It is also clear that Eqs.(7-8) can be used for arbitrary value of $\zeta$.
In the case, $\zeta =1$ one recovers immediately the WARF.
In conclusion, Eqs. (7-8) can be used to interpolate continuously 
between the WARF and the DHMRF.

Note that $b$ in Eq.(3) can be seen as the scaling factor of a massless
Gaussian field under a change of scale $2^{\zeta}$.
This remark can be understood better by noting that
the continuous (i.e. unregularized) version\cite{missarov} of the massless
Gaussian DHM is invariant under certain scale transformations. 
However, it is a $\it discrete$ scale invariance.
It seems plausible that discrete scale invariance allows
log-periodic corrections\cite{saleur} to the scaling laws.
Such a corrections were indeed observed\cite{prl} very clearly in the case of DHM.
Or goal is to find a formulation where the discrete scale invariance 
(and the unphysical features associated with it) would be replaced by
a continuous scale invariance.

The critical behavior associated with the general recursion formula (7-8)
can be studied with the usual methods.
One starts with an initial function
$Q_0(x)=rx^2+gx^4$
and for fixed $g$ determines the critical value $r_c$
by observing the transition in the behavior of $\rho _l$ defined above.
Near this critical value, one obtains the linearized expression 
\begin{equation}
Q_l(x)=Q_c(x)+(r-r_c)\lambda ^l R_c(x)
\end{equation}
where $\lambda$ is the largest eigenvalue of the renormalization 
group transformation. From this, one finds the
critical exponent
\begin{equation}
\gamma = {ln(ab^2)\over ln(\lambda)}
\end{equation}
The numerical values for various values of $\zeta $
are displayed in Fig. 1. The results indicate that $\gamma $ is a continuous
function of $\zeta $.
As $\zeta $ becomes smaller, a larger number of iterations is necessary
in order 
to obtain the critical value of $r$ with an acceptable accuracy.
For practical purposes, the number of iterations 
is of the order of $15/\zeta $. 
The limit $\zeta \rightarrow 0$ is thus difficult to reach computationally,
however we have found no indication of a drastic change of behavior 
(e.g., a sudden drop to 1) in this limit. 

We want to modify Eqs.(7-8) in order to 
get $\zeta $-independent physical quantities. In the following,
we focus the discussion on the linear behavior given by Eq. (9).
First, the $\zeta$-independence of $\gamma $ requires that
$\lambda (\zeta)=(\lambda(1))^{\zeta}$.
In addition, the renormalized coupling constants should also
be $\zeta $-independent. Defining them with the  
procedure of Ref.\cite{wilson2} (which can be 
extended straightforwardly for arbitrary $a$ and $b$), we see that
the functions $Q_c$ and $R_c$  should also be  $\zeta $-independent.
This requirement gives useful information concerning the corrections 
that need to be made to Eqs. (7-8). For instance if $Q_c$ is a 
$\zeta $-independent fixed point of such a corrected formula, one can 
set $\zeta=\zeta_0+\delta$ in the known part of the recursion formula.
Expanding in $\delta$, one can construct, order by
order in $\delta $, ``counterterms'' that cancel the 
$\zeta$-dependence. For instance at first order in $\delta$ we need to add
corrections which have a form  similar to the r.h.s of (8) but with 
insertions of $Q_l$ and its first derivative with easily calculable 
coefficients. The effects of these corrections will be investigated using 
numerical methods. We expect that this procedure can be used to
systematically improve the WARF and applied to realistic  
calculations of the critical exponents and the triviality bounds.


\centerline{\bf Figure Captions}
\noindent
Fig. 1: The critical exponent $\gamma $ as a function of $\zeta $ for $D=3$.

\vfil
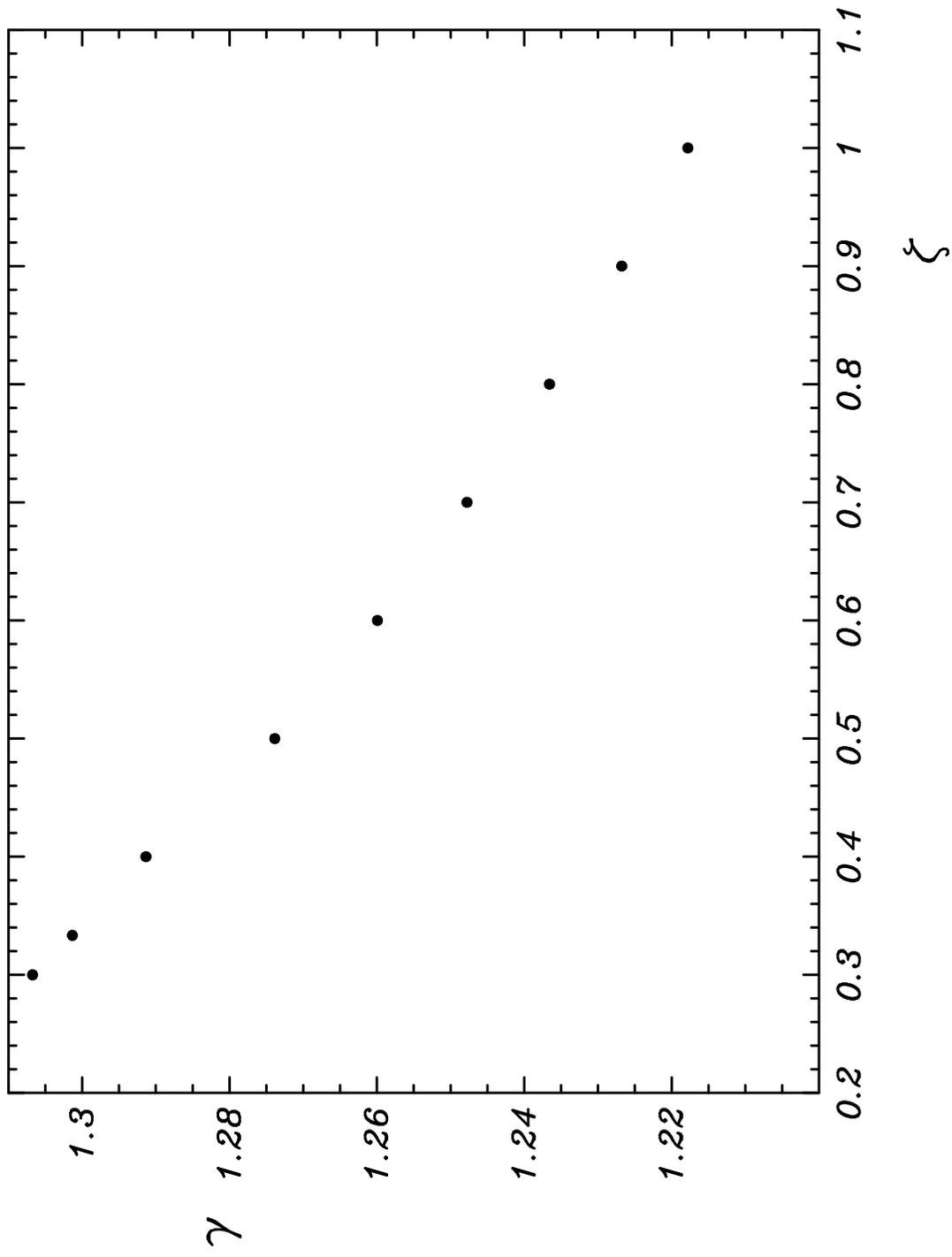
\end{document}